\documentclass[draft,wrr]{AGUTeX}

\usepackage{amssymb}
\usepackage{array}
\usepackage{amsmath}
\usepackage{bm}
\usepackage{float}
\usepackage{graphicx}
\usepackage{graphics}
\usepackage{epstopdf}
\usepackage{color}
\usepackage{xfrac}
\usepackage{mathrsfs}

\usepackage{lineno}

\authorrunninghead{ENGDAHL ET AL}

\titlerunninghead{ACCELERATING REACTIVE PARTICLE SIMULATIONS}

\begin{document}

%
%

\title{Accelerating and parallelizing Lagrangian simulations of mixing-limited reactive transport}

%
%

%
%



\authors{Nicholas B. Engdahl,\altaffilmark{1} Michael J. Schmidt,\altaffilmark{2,3} and David A. Benson\altaffilmark{2}}
\altaffiltext{1}{Civil and Environmental Engineering, Washington State University, Pullman, WA}
\altaffiltext{2}{Hydrologic Science and Engineering, Colorado School of Mines, Golden, CO}
\altaffiltext{3}{Applied Mathematics and Statistics, Colorado School of Mines, Golden, CO}

%
%

\begin{abstract}
Recent advances in random-walk particle-tracking have enabled direct simulation of mixing and reactions by allowing the particles to interact with each other using a multi-point mass transfer scheme. The mass transfer scheme allows separation of mixing and spreading processes, among other advantages, but it is computationally expensive because its speed depends on the number of interacting particle pairs. This note explores methods for relieving the computational bottleneck caused by the mass transfer step, and we use these algorithms to develop a new parallel, interacting particle model. The new model is a combination of a sparse search algorithm and a novel domain-decomposition scheme, both of which offer significant speedup relative to the reference case--even when they are executed serially. We combine the strengths of these methods to create a parallel particle scheme that is highly accurate and efficient with run times that scale as $1 / P$ for a fixed number of particles, where $P$ is the number of computational cores (equivalently, sub-domains, in this work) being used. The new parallel model is a significant advance because it enables efficient simulation of large particle ensembles that are needed for environmental simulations, and also because it can naturally pair with parallel geochemical solvers to create a practical Lagrangian tool for simulating mixing and reactions in complex chemical systems.
\end{abstract}

%
%

%

\begin{article}

%
%
%

\section{Introduction}
\label{sec:intro}
Applications of particle-tracking were historically limited to passive transport or locally first-order ({\it i.e.,} decay) reactions for many years because the particles behave independently in those systems. A clear limitation of independent particles is that it negates the possibility of any interactions between particles, including mixing-limited and multi-component reactions. This restriction has been relaxed recently, and the capabilities of Lagrangian methods for simulating mixing-limited reactive transport have been advancing rapidly \citep{Bolster2015,Benson2016,Engdahl2017,Sole-Mari2017}. Random walk particle tracking (RWPT) methods are now at the point where they are comparable to Eulerian reactive transport codes in terms of the processes they can represent. However, the RWPT methods can also add more realism to a simulation in ways like separating mixing and spreading processes and explicitly resolving small scale chemical heterogeneity or incomplete mixing, among other benefits \citep{Benson2008,Paster2015,Schmidt2018}. The cost of the increased accuracy and realism is a high computational demand, relative to classical particle tracking. A single particle in an ensemble of $N$ particles has $N-1$ possible pairs, and every relevant interaction must be evaluated at each time step. These interactions make reactive RWPT simulations difficult to accelerate because, unlike their grid-based counterparts, the nodal positions are constantly changing. For these reasons, the particle interaction step is slower than all of the other steps of the algorithm (advection, dispersion, random diffusion, and, in very simple cases, reactions) combined. Current reactive particle methods appear to have comparable run times to their Eulerian equivalents \citep{Benson2017}, but, to date, attention has focused on the accuracy of these new methods, with efficiency and speed as secondary concerns. However, speed, efficiency, and accuracy are key factors in the widespread adoption of any numerical method, and our aim is to improve all three for particle simulations.

The purpose of this note is to develop faster reactive particle simulations by relaxing the computational bottleneck caused by the mass-transfer (interaction) step. Much of the algorithm we use is unchanged from classical RWPT implementations; all the particles are simulated independently for the advection, diffusion, and hydrodynamic dispersion portions of the RWPT, as in any classical random walk. However, we consider two broad classes of acceleration schemes to speed up the added mass transfer step ({\it i.e.,} particle interactions). The first type is an efficient, sparse search algorithm known as a ``KD-tree," and the second is a simple domain decomposition scheme, novel for particle tracking applications. We evaluate 1) the accuracy of each scheme for simulating a mixing-limited, equilibrium, irreversible reaction, and 2) the speedup of each method relative to a reference ``brute force" full-matrix calculation for different particle ensemble sizes. The results show that, under serial execution, both the sparse search and the domain decomposition schemes provide significant speedup relative to the full-matrix approach, but that the domain decomposition tends to be more accurate because of slight differences in each workflow. We then develop parallelized versions of these methods and find that a combined approach using domain decomposition with a sparse search provides good scaling behavior for up to $10^{6}$ particles and $10^{3}$ processors, the largest computational parameters tested here. Overall, we find that large performance gains can be achieved with minimal computational overhead and minimal program modifications, without sacrificing accuracy. These advances show that interacting particle simulations are no longer confined to academic exercises and that they are now practical candidates for large-scale, large particle number, reactive transport simulations. Examples of the Fortran and Matlab codes used in this note are shared openly on GitHub (\textit{see} \texttt{http://doi.org/10.5281/zenodo.1476680}, \citet{DDC_repo}).

\subsection{Interacting particle models}
\label{sec:numerics}
The RWPT algorithms used herein have been described in several recent papers, so we include only a short description. Conceptually, each particle represents a finite parcel of water moving through a domain and may carry with it an arbitrary number of chemical species. Obviously, these parcels are not isolated in reality, so the particles interact with each other based on their proximity and exchange masses during each time step; readers are directed to \citet{Benson2016}, \citet{Sole-Mari2017}, and \citet{Engdahl2017} for more details.

For solute transport in porous media in $d$ spatial dimensions, the dispersion tensor is typically written $\mathbf{D} = (D_{m}+\alpha_T|\mathbf{v}|)\mathbf{I} + (\alpha_L-\alpha_T)\mathbf{v}\mathbf{v}^T/|\mathbf{v}|$, where $D_{m}$ is a molecular diffusion coefficient, $\alpha_T \le \alpha_L$ are transverse and longitudinal dispersivities, and $\mathbf{v}$ is a column vector of local mean velocity.  Many researchers have emphasized that the first, isotropic term on the right-hand side ($D_{m}$) represents mixing processes, while the next, anisotropic, term represents a hydrodynamic spreading process \citep{Gelhar1979,Gelhar_1983,Cirpka_1999,Werth_focus}; the former is a fundamental process, whereas the latter arises from having an incomplete, upscaled (volume-averaged) representation of the pore-scale velocity distribution. This conceptualization allows a natural way to partition between the particle mass transfer, which simulates true mixing, and the random walks, which predominantly simulate the spreading of particles away from each other.
Accordingly, employing this partitioning to our model and following \citet{Laboll1996} and \citet{Salamon2006}, our particle tracking algorithm begins with a simple, forward Euler evaluation of the advection and random hydrodynamic dispersion
steps for each of $N$ particles:
\begin{linenomath}
\begin{align}  \label{eq:RW3d}
{\bf x}(t+\Delta t) = {\bf x}(t) + \mathbf{A}({\bf x}(t),t) \Delta t + \left[2(\alpha_{L}-\alpha_T) \mathbf{v}_{\mathbf{x}(t)}\mathbf{v}^T_{\mathbf{x}(t)}/|\mathbf{v}_{\mathbf{x}(t)}| \Delta t \right]^{1/2} \Delta {\bf W},
\end{align}
\end{linenomath}
where ${\bf x}$ is an $N \times d$ ($d = 1, 2, 3$) vector of particle positions, $\Delta t$ is the length of a time step, $\mathbf{A}$ is a drift term that includes $\mathbf{v}$ and the hydrodynamic dispersion gradients, $\Delta {\bf W}$ is a $d$-dimensional vector of independent standard normal random variables, and $\mathbf{v}_{\mathbf{x}(t)} := \mathbf{v}(x,t)$ is the spatially variable, transient velocity field.  In 1-$d$, $\alpha_T=0$, and for incompressible flow (both of which we assume here for simplicity) $\mathbf{v}$ is a constant $v$; therefore, we have
\begin{linenomath}
\begin{align}  \label{eq:RWeq}
x(t+\Delta t) = x(t) + v \Delta t + \left[2\alpha_{L} v \Delta t \right]^{1/2} \Delta W.
\end{align}
\end{linenomath}
Higher order integration schemes, like a 2nd-order stochastic Runge-Kutta \cite[e.g.,][]{Honeycutt1992}, can be substituted for the forward Euler step to improve accuracy; an example of this can be found in \citet{Engdahl2018}.

The particle interaction scheme uses the colocation probability between particle pairs for the weights in the mass transfer process. If all particles have the same diffusion coefficient for a given species, the 1-$d$ colocation probability density function, $\nu$, is:
\begin{linenomath}
\begin{align}  \label{eq:CLD}
\nu(s_{ij}| \Delta t) = \frac{1}{\sqrt{8 \pi D_{m} \Delta t}} \exp \left[ -\frac{s_{ij}^{2}}{8D_{m}\Delta t} \right] ds,
\end{align}
\end{linenomath}
where $s_{ij}$ is the separation distance between particles $i$ and $j$, $ds$ is the representative volume of each particle \citep[see][]{Benson2016}, and the function is conditional to the time step of the random walk, $\Delta t$. Note that this approach is based on the physics of particle/particle interaction and does not adjust the kernel size to the regional concentration statistics or desired ensemble mixing and/or reaction rates \citep[e.g.][]{Rahbaralam2015,Sole-Mari2017}.
The probabilities can be written in matrix form (${\boldsymbol \nu}_{ij} := \nu(s_{ij})$), which allows us to enforce a total probability of mass transfer of 1 (mass either moves to other particles or stays put), by adjusting $ds$ so for each $i$, $\sum_j {\boldsymbol \nu}_{ij} = 1$.  We call the matrix containing these normalized probabilities $\mathbf{P}$ with entries $\mathbf{P}_{ij}:= {\boldsymbol \nu}_{ij}/\sum_j {\boldsymbol \nu}_{ij}$.

In a system with $S$ different chemical species, for any one of those species, the system of equations expressing the changes in mass between all the particles is
\begin{linenomath}
\begin{align}  \label{eq:dMassB}
dM_{i}=\frac{1}{2}\sum_{j = 1}^{N} \mathbf{P}_{ij} \left(M_{j}-M_{i}\right), \quad i=1, \dots,N,
\end{align}
\end{linenomath}
where $dM_{i}$ is the total change in mass of the $i$th particle, conditional to $\Delta t$, and $j$ is an index over the $N$ particles that are near enough to have non-negligible interactions with particle $i$. An alternative matrix form of this system is
\begin{linenomath}
\begin{align}  \label{eq:dMass}
{\bf dM} = \boldsymbol{\mathcal{P}} {\bf M},
\end{align}
\end{linenomath}
where $\boldsymbol{\mathcal{P}} := 1 / 2 \Big[\mathop{\text{diag}}(\bf P \times \bf 1 ) - \bf P\Big]$, $\bf 1$ is an $N \times 1$ vector of ones, and $\mathop{\text{diag}}(\bf z)$ is defined to be an $N \times N$ matrix with the entries of $\bf z$ on its main diagonal. Also, ${\bf M}$ is an $N \times 1$ vector of particle masses for a given chemical component, and ${\bf dM}$ is the vector holding the net change in mass to the mass vector ${\bf M}$.

The updated species/component masses due to mass transfer are then found from vector addition:
\begin{linenomath}
\begin{align}  \label{eq:NewMass}
{\bf M} (t+\Delta t) = {\bf M}(t) + {\bf dM}.
\end{align}
\end{linenomath}
The simplest, full-matrix implementation of this system requires calculating the distances between all particle pairs. The CPU time it takes a computer to complete this will scale proportionate to $N^{2}$, and the time increases linearly with the number of species/components in the simulation.

\citet{Schmidt2018} recently considered a variety of methods for simulating the mass transfer as explicit (forward Euler), implicit (backward Euler), or mixed schemes, and interested readers are referred to that article for more details. Our reference, ``brute force" mass transfer scheme is identical to the ``matrix explicit'' form in \citet{Schmidt2018}, where the masses and positions from the beginning of the time step are used to populate $\boldsymbol{\mathcal{P}}$ and ${\bf dM}$ for direct evaluation of \eqref{eq:dMass} and \eqref{eq:NewMass}. The matrix explicit method was also found to be the most accurate of all the schemes tested, though run time was not considered. Additionally, there are some obvious inefficiencies in the brute force approach since many particles can be too far apart to have non-negligible mass transfer interactions.

\section{Improved mass transfer schemes}
\subsection{Sparse search implementation}
\label{sec:spsearch}
The first class of acceleration schemes uses a sparse search algorithm referred to as a ``KD-tree," first developed by \citet{Bentley1975} and also described in \citet{Ding2013} in the context of particle tracking. This algorithm is the core of what we will call the ``looping sparse search" (LSS) acceleration scheme. This scheme was also considered in the comparative analysis of \citet{Schmidt2018}, which they called ``explicit sequential,'' and similar implementations were used in \citet{Bolster2015,Benson2016,Benson2017}.

The conceptual backbone of the KD-tree is that a binary search tree is constructed by partitioning the K-dimensional space in which the particles lie. In 1-$d$, particles above a cutoff point (often the mean or median of points being considered) are assigned to one ``branch," and particles below that point to a different branch. This branching subdivision continues to user-specified limits until only the terminal ``leaves'' remain. These leaves contain the locations of particles and high-level meta-information about the tree structure, and any higher-level nodes on the tree can be re-constructed as unions of the leaves. The type of fixed-radius searches (KD-trees are also highly efficient for nearest-neighbor searches) employed in RWPT algorithms are performed by ``climbing'' this KD-tree from the leaves upward to higher-level nodes. These searches are highly computationally efficient because most searches of a sufficiently small neighborhood do not require much climbing of the tree. As well, there are computationally-inexpensive checks that can be performed prior to searching a nearby node/leaf, and only upon a successful check would the algorithm perform the more expensive particle distance calculations. For a discussion of these computational details, see \citet{kennel}.

The original KD-tree implementation in the LSS mass transfer scheme directly evaluates \eqref{eq:dMassB} in a sequential, non-iterative manner. First, the KD-tree returns a master list of the $N$ ``parent" particles and each element in that master list contains a sub-list of the ``child" particles within some maximum separation (fixed radius), ${\bf s}_{max}$, of each particle and their separation distance to the parent. Next, the mass transfer algorithm loops through this list in a randomized order and applies the change in mass to the parent and child particles as they are encountered \citep[see][]{Benson2016}. Each child-parent pair is evaluated only once, so the algorithm naturally speeds up as it works its way through the master list of parents. The order of this looping must be randomized to avoid numerical artifacts. This means that LSS is a stochastic mass transfer scheme because over a single time step a slightly different result is obtained if a different, randomized looping order is used. The performance of fixed-radius search algorithms employing a KD-tree is proportionate to $N \log(N)$, which can be orders of magnitude faster than the brute force approach, which scales as $N^{2}$.

\subsection{Domain decomposition for particles}
\label{sec:decomp}
The KD-tree based approach is efficient because, for a properly chosen value of ${\bf s}_{max}$, most of the entries of $\boldsymbol{\mathcal{P}}$ (and by extension ${\bf dM}$) in a reasonably dispersed plume are zero because many particles are too far apart to interact. Here, we introduce a new method to accomplish something similar that also opens the door for parallelization.

We propose a simple form of domain decomposition (abbreviated DDC) where particles are grouped into discrete bins based on similar spatial positions, and each bin is evaluated independently of the others.
Conceptually, this may seem similar to the KD-tree, but this DDC generates a single collection of sub-domains rather than the bifurcating branches of a tree.
The efficiency of this approach can be demonstrated by considering the time required to build the pairwise distance matrix ${\bf s}$ (where ${\bf s}_{ij} := \vert x_i - x_j \vert$) for an ensemble of $N$ particles in serial. Assuming that the time to build {\bf s} is linearly proportionate to $N^{2}$, it is straightforward to show that building ${\bf s}$ for two populations of $N/2$ each will take half as long, since the total build time is proportionate to $2(N/2)^{2}$.
As long as each sub-domain has approximately the same number of particles within it, the theoretical speedup will be proportionate to $N^2 / P$, where $P$ is the number of ``sub-domains" the particles are grouped into, assuming a balanced distribution of particles in each sub-domain.

Domain decomposition is widely used in high-performance computing applications, but to accurately simulate the physical processes being modeled, each of the independent sub-domains must interact with the others. A ``mixed" diffusion scheme \citep[see][]{Engdahl2017} can use a combination of deterministic mass-transfer diffusion and Brownian motion where each type of diffusion applies part of the total molecular diffusion, plus hydrodynamic dispersion tensor, $\mathbf{D}$. Thus, the sub-domains might already interact any time a particle jumps randomly from one sub-domain into another. However, any time a particle's position density function overlaps a sub-domain boundary, mass transfers that should be happening are not conducted, so mixing is not simulated correctly near the sub-domain discontinuity.

A common remedy to this employed in Eulerian models is to use ``ghost nodes" that overlap and bridge the sub-domains. This typically requires an iteration over the overlapping nodes to enforce continuity, but we accomplish something similar using DDC without iterating. For a given sub-domain, $\Omega_{i}$, particles in the neighboring sub-domains $\Omega_{i+1}$ and $\Omega_{i-1}$ are included in the {\bf s}, $\boldsymbol{\mathcal{P}}$, and {\bf M} matrices for $\Omega_{i}$. Particles borrowed from $\Omega_{i \pm 1}$ are termed ``ghost particles" by analogy to their Eulerian ghost-node counterparts.
Mass transfers within a given sub-domain, $\Omega_i$, are computed by including the ghost particles for $\Omega_i$ in addition to the particles strictly within that sub-domain.
However, only the mass values belonging to particles strictly within $\Omega_{i}$ are updated during the computations on that sub-domain.

Any changes to the ghost nodes are discarded because their values will be updated in the $\Omega_{i \pm 1}$ sub-domains. This shifts the implied zero-flux boundary outside of $\Omega_{i}$, diminishing its effect. The size of the ghost particle pads, $\Delta \zeta$, is defined to be proportionate to the standard deviation ($\sigma$) of the particle colocation density, $\nu$. For our 1-$d$ example (see Eq. \ref{eq:CLD}), this gives $\Delta \zeta=\kappa\sqrt{8 D_{m}\Delta t}=\kappa \sigma$; hereafter, we choose $\kappa=3$ since this captures $99.7\%$ of the density of the total range of values in $\mathbf{P}$. The increased accuracy comes at the price of computational efficiency, which will decrease as $\kappa$ increases because more ghost particles are included in each $\Omega_{i}$.  However, the additional mass transfer calculations will reduce the actual speedup below the theoretical limit, and the reduction in speedup should be related to the ratio of active to ghost particles in each sub-domain.

\section{Test problem setup}
\label{sec:CmpAcc}
Our aim is to investigate the acceleration of the mass transfer step, so we adopt a zero-velocity system throughout this note for simplicity and only simulate diffusion ($D=D_{m}$, $\alpha_{L}=0$). The 1-$d$ domain starts at the origin and spans up to a dimensionless length $L=50$ with zero-flux (reflecting) external boundaries. A uniform diffusion coefficient of $D=1 \times 10^{0} \: [L^{2}/T]$ is defined, and diffusion is simulated as a combination of mass transfer and a random-walk, where each operation applies $D/2$ during every time step \citep[see][]{Engdahl2017}. All simulations go up to time $t=10$ using a constant time step of $\Delta t = 0.1$; this interval avoids boundary effects, and the concentration profile at the end is used for all error calculations. The initial positions of the particles are assigned from a random uniform distribution between the domain limits, and particles that diffuse (random-walk) outside these limits are reflected back into the domain. Particle ensembles of different sizes are considered, which range from 500 to 25,000 particles in this section. Because the random walk portion of diffusion and the LSS scheme are stochastic, we simulated multiple realizations to average the variability in concentrations and run times, and 75 realizations were sufficient for convergence. An example code that can be used to reproduce these serial results is available at \texttt{http://doi.org/10.5281/zenodo.1476680}, \cite{DDC_repo}.

The initial concentrations on the particles create a segregated field where two reactants, $A$ and $B$, each occupy half of the domain with unit concentration. We assume an irreversible, equilibrium reaction of $A+B\rightarrow E$, where $E$ is the mobile product. The system is simulated as two conservative components $C_{1}=A+E$ and $C_{2}=B+E$, so that at all times $E=\min\left(C_{1},C_{2}\right)$. A simple way to evaluate the error in the simulation is to compare the profile of the diffusing conservative interface to the analytical solution of the initial value problem: $C_{1}(x,t)=\frac{1}{2}\mathop{\textrm{erfc}}\left[{(-(x - x_{0}))^{2}}/{4Dt} \right]$, where $x$ is position, $x_{0}=L/2$, and $t$ is elapsed time. The analytical solution for $C_{2}$ merely substitutes $x\rightarrow -x$, but we use only $C_{1}$ since these are symmetric. The error metric for each scenario, realization, and particle ensemble size variation is the root-mean-squared error (RMSE), where the final particle positions are used for $x$ in the analytical solution.

\subsection{Hardware configuration}
The benchmark problems in this section were run on a Linux workstation (Ubuntu, version 16.04.4 LTS) with two 3.6 GHz 10-core Intel Xeon E5-2690 v2 processors. The reactive transport algorithms were written in the Matlab programming environment (version R2018a), and each random realization was run sequentially with only background load on the CPU. An important performance note is that the processors and Matlab both automatically use multi-threading by default. Multi-threading is standard in most contemporary computing systems and programming languages, so the code is not strictly serial execution, but no explicit parallelism was written into the code at this point.

\section{Domain decomposition verification}
\label{sec:dcmpscale}
DDC has not been applied to interacting particle simulations before, so we begin by investigating its accuracy and scaling. The comparison uses simulations ranging from 500 to 5000 particles ($N$) and varies the number of sub-domains ($P$) from 1 (full matrix explicit or brute force) up to 12. The relative speedup (RS) is defined as the ratio of the run time for each decomposition level divided by the run time for the single-domain simulations. Any RS value greater than 1 indicates that the domain decomposition is faster than the full matrix version, and linear scaling of RS would follow a 1:1 line (two sub-domains should be twice as fast as one, and so on).

\begin{figure}[b]
\includegraphics[width=1.0\textwidth, keepaspectratio]{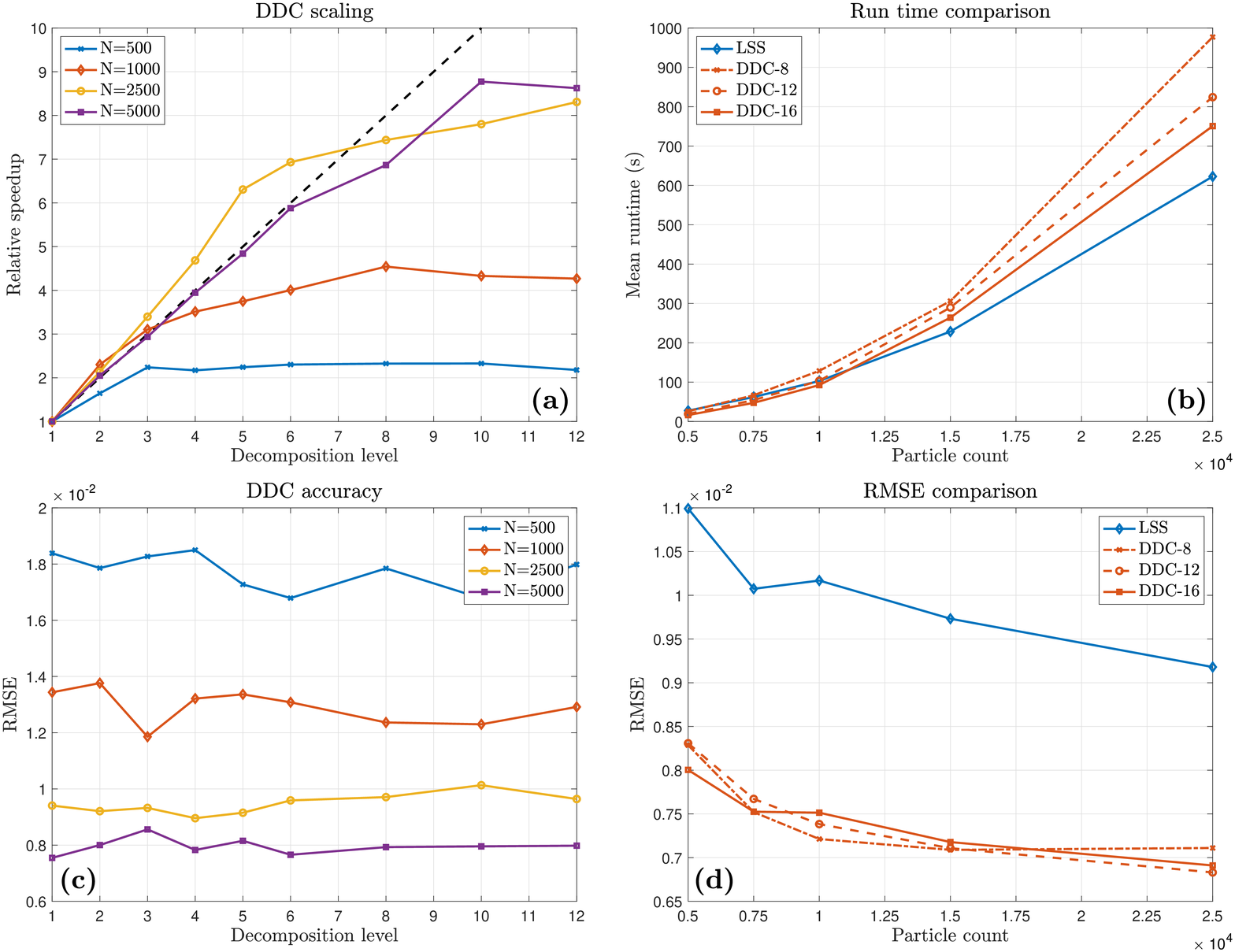}
\caption{DDC Relative speedup (a), DDC accuracy (c), LSS and DDC run time for increasing $N$ (b), and LSS and DDC accuracy (d); DDC-$P$ denotes the use of $P$ sub-domains in (b) and (d). Speedup (a) is the run time relative to the full-matrix version (DDC-1). The error is insensitive to decomposition level (c) and decreases as $N$ increases (c) and (d), which is a typical convergence pattern. However, there are no clear trends in error as the decomposition level is increased (c), indicating that DDC does not affect overall accuracy when done properly. Overall, LSS tends to have lower run times (b) but at the cost of some accuracy (d).}
 \label{fig:Fig1}
\end{figure}

The RS for this problem is shown in Fig.\ \ref{fig:Fig1}a as a function of the decomposition level (number of sub-domains, $P$); the black, dashed line indicates 1:1 scaling, and the different colors/markers differentiate the different particle ensemble sizes, $N$. Scaling along 1:1 indicates that, for example, the problem runs twice as fast with two processors as it did for one, which is considered ideal; being above this line implies scaling better than $1/P$. The speedup improves as $N$ grows and we see linear speedup for $N\ge 2500$, but this falls off for $P > 6$. At this point, the size of the sub-domain pads becomes large relative to the size of the sub-domains. Here, $D$ and $\Delta t$ give $\Delta \zeta \approx 1.34 [L]$ for the $3 \sigma$ pad. With 8 sub-domains, the lengths are $L_{\Omega}=L/8=6.25 [L]$, so $\Delta \zeta / L_{\Omega} \approx 0.21$. In other words, the combined size of the domain pads, one on each side of every sub-domain, is $42\%$ of the sub-domain's size, so a large fraction of particles in each sub-domain are non-updating duplicates. For $P = 6$ sub-domains, $\Delta \zeta / L_{\Omega} \approx 0.16$ and good scaling is still exhibited, so we suggest $\Delta \zeta / L_{\Omega} < 0.2$ as a general rule to maximize scaling efficiency.

Another important question is whether or not DDC impacts accuracy. Fig.\ \ref{fig:Fig1}c shows the RMSE for each $N$ and decomposition level, and every curve is basically flat. This shows that a $\Delta \zeta = 3 \sigma$ ghost particle pad is sufficient to prevent the DDC level from significantly impacting the accuracy. Note that the error was found to increase when a smaller pad ($\Delta \zeta = 2 \sigma$) was used, but there was no appreciable change for larger pads. A final observation is that the RMSE decreases with increasing $N$, but this is typical convergence behavior.

\section{Comparison under serial execution}
\label{sec:fullcmp}
We now briefly compare the scaling and accuracy of the \citet{Ding2013} looping sparse search (LSS) and domain decomposition (DDC) methods. The same problem described at the beginning of Section \ref{sec:CmpAcc} is simulated for both schemes. An additional 30 realizations were run for decomposition levels of 8, 12, and 16, denoted as DDC-8, DDC-12, and DDC-16, with particle numbers of 5k, 7.5k, 10k, 15k, and 25k, and the same accuracy and run time comparisons were evaluated, with the mean values of RMSE and run time shown in Figure \ref{fig:Fig1} (b and d).

Above 10k particles, LSS overtakes DDC in speed (Figure \ref{fig:Fig1}b); however, the RMSE (Fig.\ \ref{fig:Fig1}d) shows consistently higher error for LSS (about 33\% higher). There is also a clear decrease in the  run times from 8 to 16 sub-domains for 25,000 particles (Fig.\ \ref{fig:Fig1}b). For $P > 16$, DDC showed the same behaviors as DDC-10 and DDC-12 for $N=$ 5,000 in Fig.\ \ref{fig:Fig1}a, where speedup plateaued, so we did not analyze larger DDC levels. However, this plateau behavior is an artifact of our fixed domain size because DDC-16 and $L=50$ clearly violates the $\Delta \zeta / L_{\Omega} < 0.2$ bound required for efficiency.

The important result here is that both the LSS and DDC schemes offer significant performance advantages over the brute force approach. The KD-tree fixed-radius search algorithm can be found in many computational libraries and will clearly outperform the DDC as the number of particles increases (Fig.\ \ref{fig:Fig1}b), so it is the logical choice for most applications. However, the price for the speed of the LSS algorithm is that it tends to be less accurate than DDC (Fig.\ \ref{fig:Fig1}d). A mixture of the two approaches may be able to address some of these limitations, which simultaneously provides us an avenue for a fully parallel implementation.

\subsection{2-d example}
\begin{figure}[b]
\includegraphics[width=9cm, keepaspectratio]{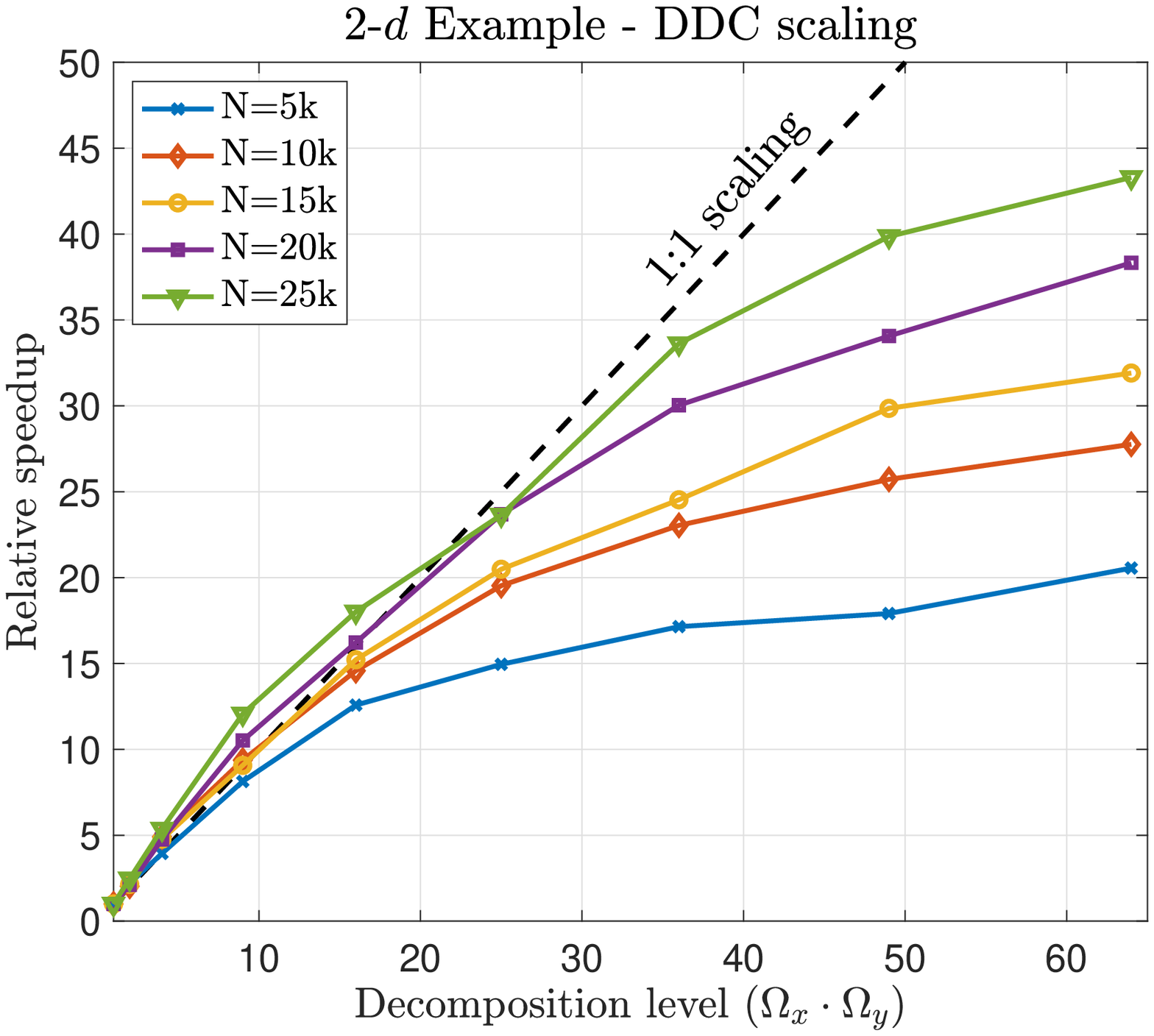}
\caption{Scaling for the 2-$d$ version of the example problem for 5k, 10k, 15k, 20k, and 25k particles ($N$). The same basic trends from the 1-$d$ case are evident here, and good speedup is achieved for all $N$ up to $P=\Omega_{x}\times \Omega_{y}=25$. Above this, the areas of the sub-domain pads are large relative to the area of the sub-domains, and the excessive, duplicated mass transfers significantly inhibit performance.}
\label{fig:Fig2}
\end{figure}

The same basic model setup can be also be used in a 2-$d$ example. Here we use a 2-$d$ domain with size $x=y=80 [L]$ with the same diffusion coefficient, time step, and total time as the previous 1-$d$ example; note that the multi-dimensional form of \eqref{eq:CLD} can be found as Eq 2 in \citet{Engdahl2017}. The 2-$d$ problem affords domain decomposition in two directions, and the way the domain is decomposed can significantly impact scaling because different approaches will yield different amounts of ghost particles. Multi-dimensional decomposition is a detailed subject beyond the scope of a technical note so we adopt a single, simple scheme for this demonstration. The total number of sub-domains is $P=\Omega_{x} \times \Omega_{y}$ where the number of sub-domains along $x$ are denoted $\Omega_{x}$, and $\Omega_{y}$ is aligned orthogonal, creating a sub-domain mesh. For all but $P=2$, we require $\Omega_{x}=\Omega_{y}$ to reduce the size of the ghost particle pads. We assume zero-flux boundaries along the domain limits in $y$ and define the initial condition as $C_{1}(x,y,t=0)=H(x)$ so the analytical solution is given by the same model as the 1D case for all $y$ values. Multiple realizations with 5k, 10k, 15k, 20k, and 25k particles were used to investigate scaling from $P=1$ up to $P=64$, and the scaling results are shown in Figure\ \ref{fig:Fig2}. As with Fig.\ \ref{fig:Fig1}c, DDC had no impact on accuracy, so the 2-$d$ accuracy plot is omitted for brevity. Significantly higher speedup was obtained for the 2-$d$ problem, and the $N =$ 25k case showed good scaling up to 36 (i.e., 6$\times$6) subdomains before falling off. Multi-dimensional scaling shows the same basic behaviors as the 1-$d$ case, where scaling efficiency decreases as the ghost particle pads increase in size. Generally, this limiting size will be similar to the $\Delta \zeta / L_{\Omega}$ criteria, but, in the 2-$d$ case, it is based on the sub-domain and ghost particle pad areas and depends heavily on how the area of the 2-$d$ domain is decomposed into smaller blocks. The main point here is that the DDC scheme is applicable and efficient in multi-domain systems, but we reiterate that the 2-$d$ decomposition scheme will be most efficient when it minimizes ghost particles, or scaling will be impacted.

\section{Parallel Implementation}
\label{sec:parallel_implementation}
Our results, so far, are for serial execution, and it should be clear that the DDC approach naturally lends itself to parallelization since the sub-domains are independent. Each sub-domain can be contained on its own processing core (with the number of cores used in a simulation defined to be $P$), and these cores only need to communicate regarding the ghost particles or when a particle ``random-walks" across a sub-domain boundary.

We consider two parallel implementations of the DDC algorithm designed to utilize the best of the LSS and DCC schemes. The first one simply executes a separate instance of the stochastic LSS (with $\textbf{s}_{max} = \Delta \zeta = 3 \sigma$) within each sub-domain, using only the particles within that sub-domain (including its ghost particles), and this method is denoted PLSS (Parallel-LSS). The main implementation detail for PLSS is that the KD-tree fixed radius search cutoff distance, $\textbf{s}_{max}$, must match the DDC ghost particle pad size, $\Delta \zeta$, in order to conserve mass, as unequal choices for these parameters will result in asymmetric mass transfers. The second parallel implementation aims to maintain the accuracy of the matrix-based mass transfer while speeding up the construction of the matrices and the linear algebra of the mass transfer step in \eqref{eq:dMass}. This is done by employing sparse matrices and sparse linear algebra techniques, and we denote this sparse parallel matrix method (SPaM). The KD-tree scheme (again, for $\textbf{s}_{max} = \Delta \zeta = 3 \sigma$) is used to build the ${\bf s}$ matrices for each sub-domain on its own core; however, even these matrices still have a large number of zeros, which causes significant computational overhead if the full matrices are stored.  As such, in SPaM, only the non-zero entries of ${\bf s}$ are stored and operated on to form $\boldsymbol{\mathcal{P}}$, which greatly reduces its size.  Mass transfers are then computed using the matrix-vector form of \eqref{eq:dMass} and are evaluated using sparse linear algebra formulas. The key distinction is that, SPaM mass transfers are deterministic (though some part of the diffusion can also be simulated as a random-walk), whereas PLSS is stochastic. The primary limitation on speedup for both of these schemes is related to the ratio $\Delta \zeta / L_\Omega$, as described in reference to the serial DDC algorithm in Section \ref{sec:dcmpscale}; otherwise, speed is increased by adding more sub-domains.

The PLSS and SPaM algorithms are implemented in Fortran using the Intel Fortran compiler version 16.0.1 and the OpenMPI message passing interface version 1.6.5. The simulations in this section solve the same 1-$d$ problem described in Section \ref{sec:CmpAcc}, and the hardware used is part of a Linux cluster running CentOS release 6.9; note that only the 1-$d$ problem is included for brevity and simplicity of the MPI implementation, but 2-$d$ scaling is expected to mimic Fig.\ \ref{fig:Fig2}. Simulations employing $P \leq 28$ were run on a 14 core (28 thread) Intel Xeon Processor E5-2680 v4. For simulations employing $P > 28$, the cluster architecture is heterogeneous; this affects consistency from run to run, but the scaling trends remain clear. Note that an example of the Matlab and parallel MPI versions of all described algorithms are included on GitHub (\textit{see} \texttt{http://doi.org/10.5281/zenodo.1476680}, \cite{DDC_repo}).

\subsection{Accuracy and run time}
\label{par_acc_time}

\begin{figure}[b]
\includegraphics[width=9cm, keepaspectratio]{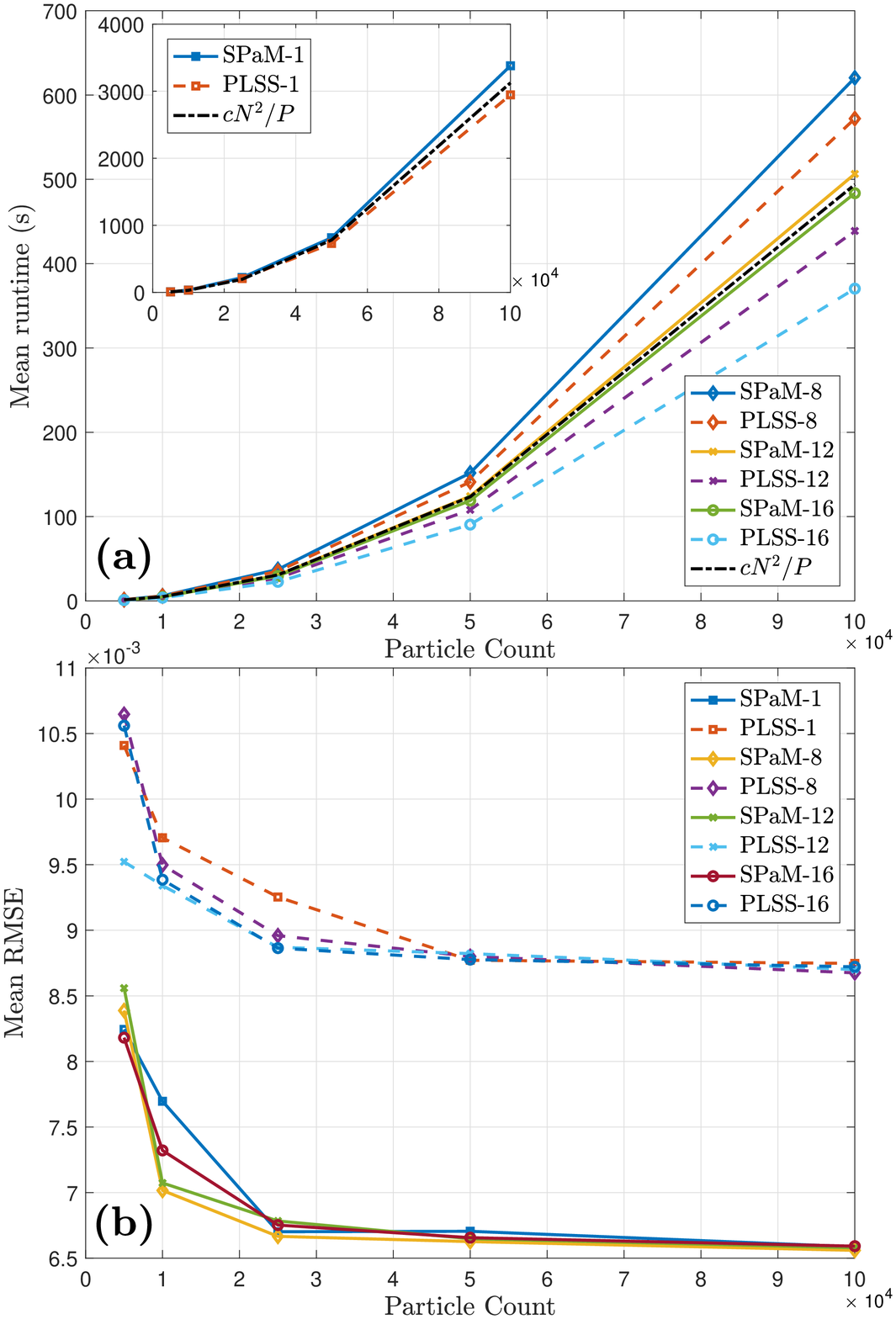}
\caption{Mean run time (upper) and mean RMSE (lower) for simulations with increasing particle number ($N$).
The two parallel domain decomposition algorithms, PLSS and SPaM, are compared for $P = \{1, 8, 12, 16\}$.
We see $\mathcal{O}\left(N^2 / P\right)$ scaling in run time with PLSS outperforming SPaM in terms of run time and the opposite result in terms of error.}
\label{fig:par_varyNp}
\end{figure}

The performance of the PLSS and SPaM algorithms was quantified in terms of run time and accuracy. 10-member ensembles of our numerical experiment were run for each algorithm with $P = \{1, 8, 12, 16\}$, where mean run time and mean RMSE for each ensemble is used for comparison. The DDC naming convention is retained, where SPaM-$P$ or PLSS-$P$, denotes that $P$ cores/sub-domains were used for that simulation.

Observing Fig.\ \ref{fig:par_varyNp}a, dramatic speedup is seen after parallelization because the run time drops from $\mathcal{O}\left(3000\right)$[s] to $\mathcal{O}\left(500\right)$[s] in the 100k particle case, relative to the serial (1-core) results (Fig.\ \ref{fig:par_varyNp}a, inset).
Both SPaM and PLSS exhibit roughly the same scaling behavior as a function of number of particles, $N$, proportionate to $\mathcal{O}\left(N^2 / P\right)$, but there is a trade-off between PLSS and SPaM in terms of run time vs.\ accuracy.
For any fixed value of $P$, PLSS outperforms SPaM in terms of run time (Fig.\ \ref{fig:par_varyNp}a); however, Fig.\ \ref{fig:par_varyNp}b, shows that SPaM displays lower error for all $P$ and $N$ combinations (though the errors are the same order of magnitude).
Finally, the PLSS and SPaM errors are insensitive to the choice of $P$, consistent with Fig.\ \ref{fig:Fig1}c.
Errors tend to decrease with $N$ (Fig.\ \ref{fig:par_varyNp}b), displaying small fluctuations with $P$; however, the fluctuations do not appear to be systematic and are likely due to the random walk portion of the solution.

The differences in run time scaling are related to the computational cost of the mass transfer step. Both PLSS and SPaM use the KD-tree to build the $\bf s$ matrix at cost $\mathcal{O}(N \log N)$ (as compared to $\mathcal{O}\left(N^2\right)$ for the brute force approach). As a result, the majority of the cost comes from the nested loop or matrix multiplication in the mass transfer step (given by \eqref{eq:dMassB} for PLSS or \eqref{eq:dMass} for SPaM). This cost is $\mathcal{O}\left(\eta N^2\right)$ for a single domain or set of particles, where $\eta$ is the fraction of nonzero entries in the mass transfer matrix (note that PLSS performs about half as many operations as SPaM in this step, hence its increased speed). The ghost particles (ideally) add little extra work to the problem, so, dividing this work among $P$ processors, the total cost of the parallel algorithms is $\mathcal{O}\left(N^2 / P\right)$.

\subsection{Parallel scaling and efficiency}
\label{par_scale}

\begin{figure}[b]
\includegraphics[width=1.0\textwidth, keepaspectratio]{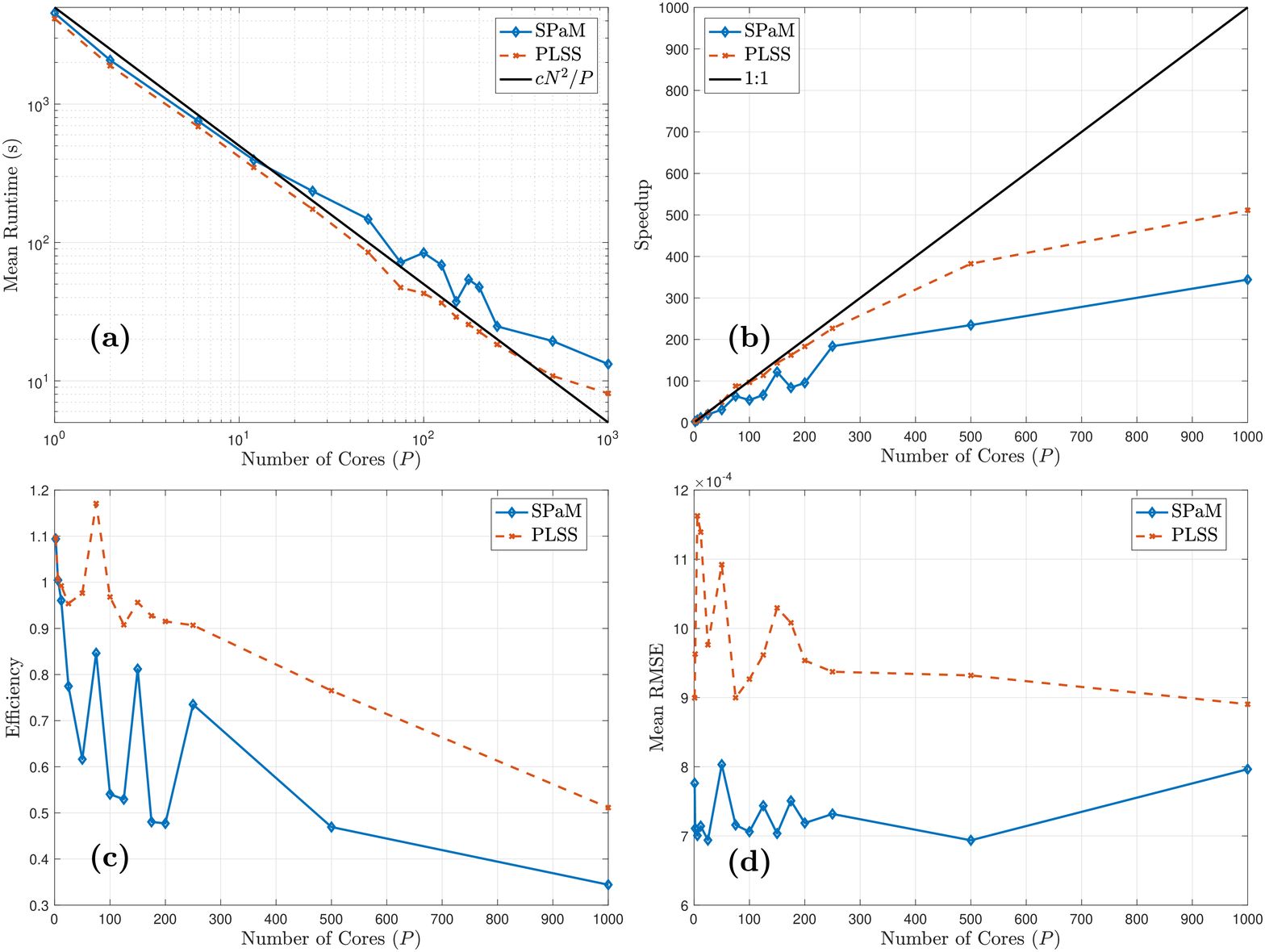}
\caption{Mean run time (upper left), speedup (upper right), parallel efficiency (lower left), and mean RMSE (lower right) for simulations with fixed particle number $N = 1 \times 10^6$.
The two parallel domain decomposition algorithms, PLSS and SPaM, under increasing $P \in [1, 1000]$. $c \approx \mathcal{O}\left(10^{-8}\right)$ is a linear pre-factor describing the trend.
We see $\mathcal{O}\left(N^2 / P\right)$ scaling in run time with PLSS outperforming SPaM in terms of run time and the opposite result in terms of error.
We also observe linear or nearly-linear speedup for $P < 500$, at which point the suggested bound of $\Delta \zeta / L_\Omega < 0.2$ is violated.
Efficiency stays high until the bound on $\Delta \zeta / L_\Omega$ is violated, and the oscillations are attributable to the heterogeneous computing architecture.}
\label{fig:par_N1e6}
\end{figure}

Finally, we investigate the parallel scaling and efficiency of the PLSS and SPaM algorithms for increasing numbers of computational cores.
The test case is similar to the that of Section \ref{sec:CmpAcc}, but the number of particles is fixed at 1 million and the domain is expanded to $L = 5000$; this keeps the ratio $\Delta \zeta / L_\Omega$ smaller than 0.2 for most of the cases tested, but $P=1000$ shows what happens when this bound is violated.
Each numerical experiment uses 3-run ensembles for $P \in [1, 1000]$, and we consider 1) the mean run time and RMSE for each simulation, and 2) the speedup and parallel efficiency for both algorithms.
Speedup ($S_P$) and efficiency ($E_P$) for a simulation employing $P$ cores are defined to be
\begin{linenomath}
\begin{align*}
    S_P &:= \frac{T_P}{T_1},\\
    E_P &:= \frac{S_P}{P},
\end{align*}
\end{linenomath}
where $T_P$ is the mean run time for a simulation employing $P$ cores.

Considering Fig.\ \ref{fig:par_N1e6}a, we see decreasing run times all the way up to 1000 cores, and we observe the same $\mathcal{O}\left(N^2 / P\right)$ scaling behavior as in Section \ref{par_acc_time}.
Looking to Fig.\ \ref{fig:par_N1e6}b, we see initially nearly linear speedup for low numbers of cores, with SPaM becoming slightly sub-linear near $P = 100$, and both algorithms strongly breaking from the linear trend for $P > 500$.
The initial break from linear speedup for SPaM is likely related to the heterogeneous computing architecture that was available for this experiment and we expect a trend closer to linear if a homogeneous architecture was used.
The second transition into permanently sub-linear speedup, however, is related to nearing and then violating the $\Delta \zeta / L_\Omega < 0.2$ suggestion, since for $P = 250$ we have $\Delta \zeta / L_\Omega \approx 0.1$, but for $P = \{500, 1000\}$, we have $\Delta \zeta / L_\Omega \approx \{0.19, 0.38\}$, respectively.
Examining Fig.\ \ref{fig:par_N1e6}c, both PLSS and SPaM exhibit greater than 100\% efficiency in some cases, likely reflecting multi-threading effects similar to the super-linear speedup seen in Fig.\ \ref{fig:Fig1}. As $P$ increases, PLSS stays above 90\% efficiency until the final break from linear speedup at $P = 500$, while the behavior of SPaM is similar but more erratic, likely due to subtle differences in memory management between the two algorithms.
The key point, however, is that the error remains insensitive to $P$ (Fig.\ \ref{fig:par_N1e6}d), and that, even when employing a large number of cores (1000), we see good efficiency for PLSS and SPaM when $\Delta \zeta / L_\Omega < 0.2$.

\section{Discussion}
\label{sec:Discusting}
The serial version of our particle method is operationally simple and generally has fewer numerical issues than most grid-based schemes, and the same is true for the parallel implementation. The only criteria for accuracy is that the ghost particle pads are sufficiently wide, and good scaling efficiency is maintained when $\Delta \zeta / L_\Omega < 0.2$. Under such conditions, the serial (DDC) and parallel schemes (PLSS and SPaM) show scaling proportionate to $\mathcal{O}(N^{2}/ P)$, (\textit{i.e.} linearly with P) for fixed $N$, which offers significant speedup compared to the brute-force method (run time proportionate to $N^{2}$). Even when the efficiency scaling begins to break down, additional speedup can still be realized for $10^{6}$ particles on $10^{3}$ cores. Some additional considerations are needed for 2- and 3-$d$ systems to efficiently define the sub-domain limits, but many of these details can be borrowed directly from adaptive meshing techniques that also add minimal computational overhead and the scaling should be expected to follow Fig.\ \ref{fig:Fig2}. The parallel architecture can also easily accommodate reactions when they are evaluated independent on each particle after the mass transfer step \citep[see][]{Benson2016,Engdahl2017}. However, the inclusion of intricate geochemical reactions may have a significant impact on performance scaling, but this is beyond our current focus. We also note that adaptive kernels \citep[e.g.][]{Rahbaralam2015,Sole-Mari2017} can be used in these acceleration schemes, but $\Delta \zeta$ must become a time dependent parameter that adjusts the sub-domain pad as the interaction kernels change size, which will impact scaling. Overall, our main point is that this note has shown that parallelized, interacting particle simulations are stable, numerically efficient, and practical tools for large particle number simulations of complex reactive transport processes.

%

\begin{acknowledgments}
This work was partially supported by the U.S. Department of Energy, Office of Science, Office of Biological and Environmental Research, under award number DE-SC0019123; the US Army Research Office under Contract/Grant number W911NF-18-1-0338; and the National Science Foundation under awards EAR-1417145, DMS-1211667, and DMS-1614586.
The online resources located on GitHub (\textit{see} \texttt{http://doi.org/10.5281/zenodo.1476680}, \citet{DDC_repo}) include the Matlab scripts and Fortran code necessary to reproduce the results.
\end{acknowledgments}





%

%
%

\end{article}
%
%
%
%
%
%
%

\end{document}